A File System For Write-Once Media

Simson L. Garfinkel and J. Spencer Love

Simson L. Garfinkel
MIT Media Lab
Cambridge, MA 02139
Garfinkel@MULTICS.MIT.EDU

J. Spencer Love
MIT Information Systems
Cambridge, MA 02139
JSLove@MULTICS.MIT.EDU

This research was performed at the MIT Media Lab during the spring and summer of 1985, and was sponsored in part by a grant from IBM.

## Abstract

A file system standard for use with write-once media such as digital compact disks is proposed. The file system is designed to work with any operating system and a variety of physical media.  Although the implementation is simple, it provides a a full-featured and high-performance alternative to conventional file systems on traditional, multiple-write media such as magnetic disks.

Note on program examples in this article:

All data structures and program examples in this document are presented in the form of C language fragments. These are living excerpts from the Media Lab's CDFS implementation. Since the definition of C's int, short int and long int differ from implementation to implementation and computer to computer, we have avoided their use in our examples.  Instead, the examples only reference derived types such as int16 and int32. These derived types are defined separately for each compiler.

Intnn defines an unsigned integer nn bits wide, stored low-byte first. Sintnn defines the corresponding signed-integer.  char foo[nn] allocates space for an array of nn 8-bit characters.

1. Introduction

A preformatted, blank compact disk[A compact disk (CD) consists of a thin sheet of metallic film covered by a transparent shield.  Information is stored in the form of non-reflective regions (holes or bumps) on a narrow spiral track in the metallic film. The CD is scanned by a solid-state laser that reflects light off the metallic film.  A significant amount of the information stored on a conventional audio-format compact disk is tracking and error correcting information. [ecc-info] A more powerful laser can record information on compact disks.  The error correcting codes make it impossible to rewrite blocks already written on the disk.]  costing under ten dollars. It can contain over 500 megabytes.  The error rate is less than 1 error per $10^{14}$ bits read.  Drives capable of reading 176,400 bytes per second [d5] have already been delivered at a retail price of less than three hundred dollars per drive.

The CD reader hardware is currently available for computers. [philips] Compact Disks used for digital data storage are commonly referred to as CDROMs, reflecting their read-only nature. Applications such as software distribution and electronic publishing are natural candidates for the CDROM technology. In

the absence of a file system standard for CDROMs, vendors have introduced
products with proprietary, read-only imitations of existing read-write file
systems. [laserdata]

   A direct read after write (DRAW) device is capiable of writing each block on
the compact disk once. The consumer demand for these devices is expected to be
considerable.  We expect that the same economies of scale that have provided
the consumer electronic market with low-cost read only devices will soon supply
the market with low-cost write-once devices.

   Existing file systems will not work with DRAW devices. Their operation is
predicated on the assumption that any block, once written, can be rewritten.
This is assumption is made fully clear when blocks containing directories and
file maps are considered. [unix-fs] [organick]

2. CDFS Design Goals
   Our overall goal is twofold: First, to be able to maintain files on a
compact disk, enjoying the benefits that he media affords.  Secondly, we want
to be able to carry our files on CD to any workstation or computer installation
and be instantly able to use and update the stored data. CDFS must provide
transportability between different sites and different operating systems: it
must be operating system independent. To accomplish these goals, the file
system must possess the following features:

   - Retain the useful properties and general appearance of a
     multiple-write, hierarchical file system.

   - Store directory information on the same write-once media as the
     contained files.

   - Minimize file system overhead, including storage space, processor
     work and seek activity.

   - Keep a complete audit trail of every modification to the media.

   - Be independent of the underlying media interface and data
     organization.

   - Never modify data previously written to the disk.

   - Permit minimum memory implementations so that file system structures
     need not remain scarce buffer memory.

   - Allow for easy, automatic reconstruction of the file system after
     media failure or interrupted update.

   - Allow for the storage of operating system specific information.

   - Make provisions for alternative file systems, either in partitions or
     embedded in files.

   - Be simple to implement.

   - Be extensible.

   For most purposes, CDFS is a functional superset of most file systems
commonly in use. That is, it implements most of the conventional abstractions
(files, directories, links, timestamps) found on conventional file systems.

This is to allow a hierarchy resident on magnetic media to be copied to CD and back to magnetic media without change to the hierarchical structure or contents.

3. Transactions
   CDFS organizes information on the write-once media in sequences called Transactions. Each Transaction consists of a group of files, the directories containing those files, a Directory List and an End-Of-Transaction block (EOT).

   If any updates are made the current Transaction must be completed prior to dismounting the media.  The last block written on the compact disk at the end of a Transaction must be an EOT. Contained in the EOT (See figure EOT-PICTURE) is a pointer to the most recently written Directory List. Each Directory List consists of associated header information (see figure DIR-LIST-HEADER) and an ordered array with one entry for each directory in the file system (see figure DIR-LIST-ARRAY).

   Each entry in the Directory List array contains a pointer to the location of the most recent version on the disk of that particular directory, which, in turn, contains pointers to the most recent versions of the individual files it contains.

   We arrived at the Directory List concept after taking a census of a number of heavily used time-sharing systems. The systems we examined had a relatively small number of directories when compared to the total amount of online information. For example, MIT-MULTICS, a Honeywell DPS-8/70M serving a user community of over 3000 researchers, has 4.26 gigabytes on line, 3.12 gigabytes of storage actually in use, but only 9567 directories. On the basis of this and other censuses, we do not expect that the directory list will require a disproportionate amount of storage space.

   The Directory List allows directories to be quickly located since it contains a direct pointer to the most current version of every directory. It additionally permits new versions of directories to be written without updating their containing directories. Thus, if in the course of a single transaction a file is added to a directory deep in the hierarchy, it is only necessary to rewrite the containing directory, the Directory List and the EOT. The cost of maintaining the directory list is much less, both in terms of storage space and processor overhead, than the alternative of complete hierarchy rewrites.

4. An Example Transaction
   Let's follow the development of a file hierarchy on a compact disk over the course of two Transactions. In this example, Transaction update occurs as in the Media Lab's CDFS implementation. Other implementations might perform transactional update differently.

   CDFS only writes blocks consecutively. In this example, the CD starts blank. In the first Transaction, the compact disk is created with two files, "life.c" and "wheel.c", in the root directory.  In the second Transaction, the file "life.c" is modified while the first file, "wheel.c" remains untouched.

   At the close of the first transaction, the two files have been written, along with the associated Directory List and EOT.  A pictorial view of first few CD blocks is presented in figure FIRST-TRANSACTION. A description of the purpose of each block follows.

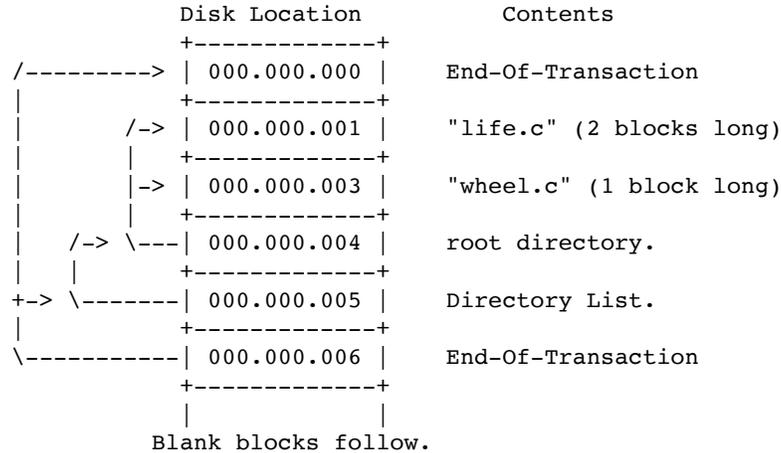

```
              Disk Location        Contents
              +-------------+
     /--------->| 000.000.000 |    End-Of-Transaction
     |        +-------------+
     |     /-> | 000.000.001 |    "life.c" (2 blocks long)
     |     |   +-------------+
     |     |-> | 000.000.003 |    "wheel.c" (1 block long)
     |     |   +-------------+
     |  /-> \---| 000.000.004 |    root directory.
     |  |      +-------------+
     +-> \------| 000.000.005 |    Directory List.
     |        +-------------+
     \---------| 000.000.006 |    End-Of-Transaction
              +-------------+
              |             |
           Blank blocks follow.
```

Figure 4-1:   A Sample transaction

Note: the notation iii.jjj.kkk denotes a particular block on the compact disk. (iii.jjj.kkk) refers to the block located at minute iii, second jjj, block kkk.

Arrows (--->) in the above diagram indicate pointers from the End-Of-Transaction block to the Directory List, from the directory list to each directory, and from each directory to the files it contains. Although there are many other pointers stored in these blocks, they have been omitted from the drawing for clarity's sake.

Block Location   Explanation

000.000.000      The first block of a CDFS generated disk contains an EOT.  EOTs
                 contain media-specific information about the disk, such as the
                 recording format, CD pointer (called "cdblock" and explained
                 later) format, and the version of the CDFS that the disk was
                 created with. This block also contains the name of the disk's
                 owner, the name of the site which created the CD and other
                 human-readable information. (See figure EOT-PICTURE for the EOT
                 structure.)

                 It is the responsibility of the Driver Layer to locate the
                 first EOT. When the CD is mounted, CDFS derives enough
                 information from the first block to locate any other block on
                 the disk.

000.000.001      The file "life.c" is stored contiguously, preceded by
                 out-of-band information stored in a File header. (See figure
                 FILEHEADER-FIGURE for the fileheader structure). The contents
                 of the file are positioned independently of the fileheader, but
                 in this example follow contiguously.

000.000.003      "wheel.c" is the second file stored on the CD, stored as a
                 fileheader followed by file contents.

000.000.004      At the close of the transaction, directories that have been
                 modified as a result of insertions, deletions or modifications
                 are rewritten. Directories consist of a fileheader, a block of
                 data pertaining to the entire directory, and an ordered array
                 of elements describing each entry in the directory. (See figure

DIRECTORY-FORMAT for the directory structure). In this example, the root directory contains two entries.

000.000.005    The Directory List follows the directories. (See figures DIR-LIST-HEADER and DIR-LIST-ARRAY for the directory list structure.)

000.000.006    The closing EOT is the last block written to the CD.

After the second Transaction, in which a new version of "life.c" is written, the pictorial view would resemble figure SECOND-TRANSACTION. Note that none of the blocks written in the first transaction have been modified.

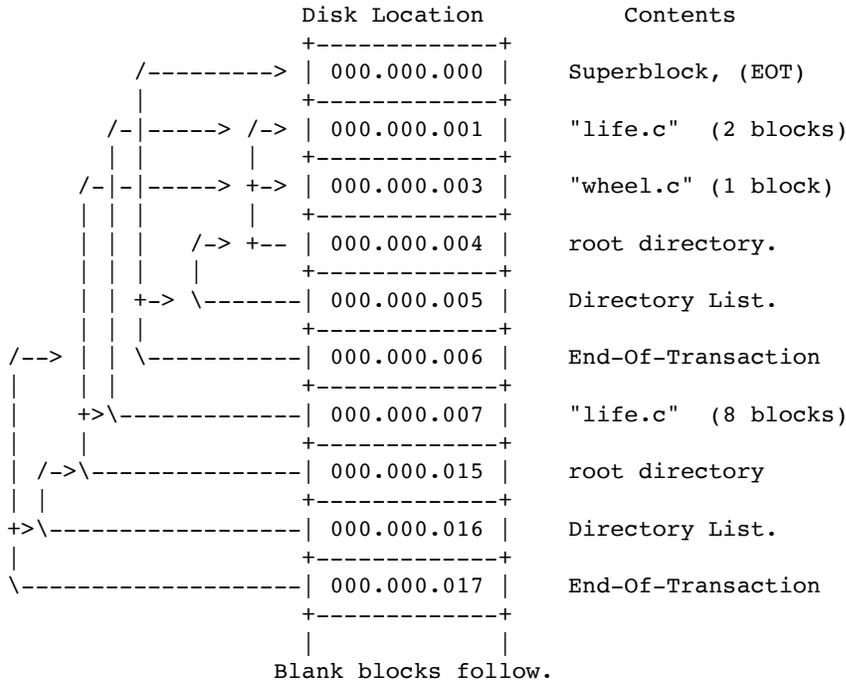

Figure 4-2:   A second sample transaction

New block      Explanation

000.000.007    The first file written on the second transaction is the updated version of "life.c". The new version of "life.c" contains a pointer to the previous version.

000.000.015    A new root directory is written out following the modified files. The directory points to the most recent version of the fileheaders of the files it contains. It additionally points to the previous version of the root directory, to allow retrieval of deleted files.

000.000.016    The Directory List is written following the last directory. It contains a pointer to the most recent version of each directory.

000.000.017    The End-Of-Transaction block is written last. It contains a pointer to the previous End-Of-Transaction and to the current Directory List.

Note: Directories, Directory Lists and End-Of-Transaction blocks may be
written at any time. The batched nature of Transactions described here is a
result of our implementation, and arises from a desire to conserve the storage
overhead associated with file update, at a cost of maintaining the contents of
modified directories and the directory list in memory until the end of the
transaction.

When the disk is mounted, CDFS reads the disk's format from the the first
block on the disk, then locates the last modified block. From the final EOT,
CDFS obtains the location of the directory list. The directory list contains
pointers to the current version of every directory on the disk. Each directory
contains pointers to the fileheaders of all non-directory contents
(subdirectories are located via the directory list). Fileheaders point to file
contents. In this manner, locating the EOT allows any byte resident in the
filesystem to be easily located.

5. Driver Layer
   The CDFS Driver Layer is responsible for reading and writing data to the
DRAW device and device control.

   Locations on the write-once media are denoted by "cdblock" pointers. A
cdblock pointer consists of 64 bits partitioned into a group of bits which
denote a block address and a group of bits which denote a byte offset in that
block. Thus, a cdblock pointer points to a particular byte on the compact disk.

   The partitioning and interpretation of the 64 bit cdblock quantity is the
responsibility of the Driver Layer. Presumably, the block address field will be
further partitioned in a way reflecting the addressing scheme used by the DRAW
hardware. Details of the partitioning scheme are recorded in the pointerdes
array located in every End-Of-Transaction block on the disk.

   In the case of audio-format CDROMs, the cdblock is partitioned into 48 bits
of block address and 16 bits of byte-offset in block. The block size is 2048
bytes. The 48 bit block address is further partitioned by the driver into three
16 bit fields: minute, second and block number. Other partitionings are
possible. When this partitioning scheme is used, the contents of the pointerdes
array in every End-Of-Transaction naturally follows as:

| Name   | modulo_of_value | bits_in_value |
|--------|-----------------|---------------|
| Minute | 70              | 16            |
| Second | 60              | 16            |
| Block  | 75              | 16            |
| Offset | 2048            | 16            |

   The remaining entries in the pointerdes array are blank and ignored.

   Note that the "Name" is not stored in the pointerdes array, but is included
for clarity's sake only. The block size is inferred from the modulo of the last
entry in the pointerdes array.

   It is the responsibility of the driver layer to provide the following
functions to the CDFS:

   - Return physical information pertaining to the disk currently mounted,
     including block size and last addressable block on the disk.

   - Read a block, returning the contents, if possible, or an error

indication.  Error indications must distinguish between a virgin
   block and an unreadable block.

- Write a block on the CD. If blocks may only be written in sequence, a
  suitable error code must be returned.

- Calculate successive cdblock addresses. (Address arithmetic depends
  on media format at the hardware level. The driver layer is
  responsible for mapping device addresses onto a simple linear address
  space.)

- Find the first virgin block within a given range. If this primitive
  is not available, CDFS will locate the first virgin block with a
  binary search.

6. Current Implementation
   The first implementation of CDFS is a C subroutine library on a Digital VAX 11/785 running Berkeley 4.2 UNIX. The subroutine library is now being modified for use with the MSDOS and VMS operating systems. Programs which wish to access files stored on a CD must be specially modified and linked with the CDFS library.

   We used a DRAW simulator to write the CDFS before DRAW devices were commonly available. The simulator, running under UNIX, made a large disk file take on the appearance of a small, write-once compact disk.  An entire 424MB disk pack was used for the simulator. The simulator-based system can double as an editing system for the development of read-only CDs: files are copied to the simulated CD.  When all of the desired files are in place, the UNIX file is copied to a tape which is used to master the CDROM.

   At first, CDFS will be used for user file storage in a public workstation environment. During a work-session, the user will copy the files he is working on from CD to a local hard disk on the workstation. At the end of the work-session, modified files would be copied back to CD. Eventually, we hope to be able to support diskless workstations and automatic copying of modified files back to CD.

   Work is underway on a file-server implementation of CDFS. In this implementation, a CD-equipped computer will be attached to a local area network. All the computers on the network will be able to keep files on the CD. There will be no difference, from the user's or programmer's point of view, between files stored on compact disk and files stored on conventional magnetic devices.

   The Media Lab CDFS implementation lacks fragmented files, addnames, resolution of links and handling of reserved properties. These features will be added as needed during the fall of 1985 as the CDFS user and programmer community increases.

6.1. Programmer's interface
   A programmer writing CDFS applications will interact with the CDFS through the Programmer's interface (currently implemented as a subroutine library written in C). The programmer's interface requires the programmer to know nothing of the actual structure of how files are stored on the CD. Files are accessed by name, directories are accessed by number. The Media Lab compact disk programmer's interface provides the following functions:

## 6.1.1. Calls for use with directories

The following functions in the programmer's are primarily for dealing with directories and their contents. Most of the calls require a directory number and an entry name.

```
CD_mkdir(int32 parent,string name)
```

Create a directory in a given parent directory. File Systems are initialized with an empty root directory. The root directory has no name. The number of the root directory is "1". This function returns the number assigned to the newly created directory.

```
CD_dir_number_to_name(int32 dirnum,string delimstr,
        string replacestr)
```

Convert a directory number to a full pathname. Directories in the pathname are separated by delimstr. Any occurrences of delimstr in the pathname, other than those being used for the purpose of delimiting directories, will be replaced by replacestr

The use of delimstr and replacestr allow full pathnames to be formatted in a manner similar to the host operating system. In a MSDOS based implementation, the backwards slash character ("\") would be used as delimstr, while the forwards slash ("/") would be used on a UNIX implementation and the greater-than sign (">") would be used on a MULTICS or Lisp Machine implementation.

```
CD_delete_entry(int32 dirnum, string entryname)
```

Detaches a file or directory from its parent. If entryname is a directory, it need not be empty. If entryname is null, the directory is deleted. (Note that the root directory may not be deleted.)

```
CD_get_direntries(int32 dirnum,string entryname,
        int32 &entry_count)
```

Returns the contents of a directory that match entryname. If entryname is null, return all of the contents. entry_count is set to the number of entries returned.

```
CD_rename_entry(int32 dirnum,
        string oldname,string newname)
```

Renames an entry in a directory and rewrites the entry's fileheader.

```
CD_undelete_entry(int32 dirnumber,string entryname,
        int32 version,bool assign_new_filenumber)
```

Reattaches a entryname to its parent directory. Undeleting version number 0

undeletes the most recent version of the file or directory with this name. The entryname must have been previously deleted (detached) with CD_delete_entry. If assign_new_filenumber is true, a new filenumber will be assigned to the undeleted file; otherwise, the original filenumber will be used.

This operation searches through previous versions of the directory for the requested entry and reenters the entry into the directory.

```
CD_name_to_file_number(string pathname, int32 context,
        int32 &containing_dir, int32 &filenumber,
        string downdir, string updir, bool updir_is_a_dir)
```

Resolves relative or absolute CD pathnames to directory number and file number. Allows the use of arbitrary down directory and up directory notations, including whether the up directory identifier is a directory (for example, the directory ".." in the UNIX operating system) or a delimiter (for example, the "<" delimiter in the Multics operating system).

```
CD_get_fileinfo(int32 dirnum, string filename,
        int32 file_version, int16 returned_structure_version)
```

Returns an info-structure on the contents of a given filename in a given directory. Specifying version "0" returns info on the most recent version. Differing return structures are provided; in particular, a version is supported which can be derived on the basis of information stored in the directory, eliminating the need for a seek to the fileheader.

```
CD_destroy_file(int32 dirnum,string cd_filename,
        int32 version)
```

Destroys all information associated with the given filename in the given directory. This operation is performed by the overwriting of previously recorded blocks. Used for destroying sensitive information. If version number is 0, all versions of the file are destroyed, otherwise only that version is destroyed. If cd_filename is blank, the directory is destroyed.

6.1.2. Calls for use with files
   The following functions are primarily for dealing with files on the CD as files. While most calls, like as those above, take as arguments a directory number and a entry name, these calls are more concerned with the side effects on the entry names themselves, rather than on the directory.

```
CD_copy_file_to_cd(string native_filename,int32 dirnum,
        string cd_filename,bool start_on_next_block,
        bool preserve)
```

Copies a file from the native file system to the CD file system. If start_on_next_block is true, the contents of the copied file start on a block boundary (important for digital music and video applications). If preserve is true, the file is copied with the same modify times and ownership as existent on the native multiple-write file system.

```
     CD_copy_file_from_cd(int32 dirnum,string cd_filename,
             string native_filename,int32 version,bool preserve)
```

   Copies a file from the CD file system to the native file system. Specifying version "0" gets the most recent version. If preserve is TRUE, then modify times, ownership and protections on the file are restored as they were on the CD, if possible.

```
     CD_fopen(int32 dirnum,string cd_filename,string mode)
```

   Opens a stream for reading or writing files on the CD. Note: Only one write stream may be open at a time.

```
     CD_fread(CDFILE *cd,string buffer,int32 nchars)
```

   Reads from a stream opened with CD_fopen.

```
     CD_fwrite(CDFILE *cd,string buffer,int32 nchars)
```

   Writes to a stream opened with CD_fopen.

```
     CD_fseek(CDFILE *cd,int32 offset,int16 relative)
```

   Position a read-stream relative to the beginning, end or current position in a CD file.

```
     CD_fclose(CDFILE *cd)
```

   Close a read or write stream.

   The Programmer's Interface currently lacks primitives for creation and deletion of links addnames, and the maintainence of property lists.

6.2. Multiple Partitions and embedded file systems
   CDFS supports the allocation of a part of the media outside the file system. There are two ways to allocate media outside of the file system; both allocation schemes can only take place at the time of the media's first use. Both schemes take advantage of the fact that the first block on the media contains an End-Of-Transaction block which is used to determine the usable blocks of the media.

   In the first preallocation scheme, the first block on the media is written so as to restrict CDFS from writing past a given block by specifying a modulo in the pointerdes array smaller than that of the physical media; this effectively restricts CD files to, for example, the first half of the disk. The remainder of the disk is available for extra-CDFS use.

   An alternative approach is to embed the preallocated space within a file

resident in the file system. This scheme makes use of the next_eot_location
field in the first EOT on the disk. If this pointer is null, then the entire
media is searched for the last end of transaction block while mounting. If it
is not null, then only the part of the media between the end of the first
transaction and the end of the media is searched.

   CDFS always writes to the first unwritten, unallocated block on the media.
One reason this is done is to facilitate the mount-time binary search for the
last-written EOT. If there were unwritten blocks in the area searched that
preceded the end of the last transaction, the binary search might find them and
become confused. Partitions may be placed wherever they can be excluded from
the search.

   If some application requires the ability to do its own media management,
arbitrary size areas of the media within the first transaction can be allocated
for this purpose. As an aid to locating them, they can appear as files within
CDFS.

   Partitions which appear as files in CDFS can be read and written using CDFS,
although any block, once written, cannot be written again.

6.3. CDROM and Electronic publishing
   Software and databases are currently being sold on CDROMs. In the future, we
expect text and reference books to be sold this way as well. The CDROM is
simply a prerecorded CD. Using a write-once drive, it will be possible to
record additional information on the CDROM. If a CDROM is sold in CDFS format,
then CDFS can be used to annotate and update the CDROM's contents. An
unmodified CDROM can be mounted without the binary search for the last
transaction, provided that the entire initial contents of the CDROM are the
first transaction, so that the first block on the media points to an end of
transaction block followed by an unwritten block.

   A book could be sold in CDROM format with a directory for each chapter, and
subdirectories for sections. The reader could enter "marginal notes" using a
DRAW drive. The directory tree would constitute an outline of the book, and
CDFS links could be used for cross references. If special software could use
the book more effectively, demonstrate points or animate examples, programs for
a dozen different operating systems and digitized images could be stored on the
CD as well.  This barely comments on the possibilities of electronic
publishing.

6.4. Garbage Collection
   As a CDFS updates media, a certain amount of garbage is generated, in old
versions of files, directories, and transactions. Eventually, the disk will
fill up. In some cases, little compaction will be possible because new files
have been created rather than updating old ones. In other cases, where CDFS
will have been used to emulate a multiple-write file system, a great deal of
compaction will be possible by copying the most recent version of every file to
a blank disk. While compacting a CD containing many megabytes of active files
will take time, it should be possible to access and even update the hierarchy
while the files are being transferred.

7. Current areas of development

7.1. Fragmented Files
   The existing CDFS implementation does not yet implement fragmented files,
but formats for them have been defined. Fragmented files permit multiple files
to be open for update concurrently. Without fragmentation, implementations must either restrict the number of open write files to one or buffer write operations on a file-by-file basis.

   Between the file header and the file's contents, fragmented files contain a file map which describes the physical location of the data contained by the file. The file_location field of the fileheader points to the file map instead of the file contents. The file map is a contiguous array of fragment descriptors which describes the mapping of valid data stored on the disk to data in a logical file.  Data is stored in fragments, or strips, of valid data, which reside in arbitrary locations on the media.

   Each fragment descriptor consists of three items: the position in the file where the fragment begins, the number of bytes in the fragment, and a cdblock pointer to the fragment, or strip, on the media. The strips may logically overlap but the data must be identical in regions of overlap.  Overlap permits retaining copies of valid data which have already been written. Not every logical byte of a fragmented file need be mapped to a physical location on the disk, but an error condition will be raised if one of these bytes is attempted to be read.

   Fragments occupy real space on the disk. Blocks containing fragments do not begin with any special header. Although the implementation should be able to read files with fragments as small as one byte, it should, for efficiency, never write fragments smaller than at least several thousand.

   Fragmented files can save space on the storage media in applications requiring minor updates to large datafiles. If one byte changed in a file a million bytes long, a new fragment of minimum size can be written out, rather than rewriting the entire file.

   A contiguous file can be rewritten as a fragmented file simply by writing out a new file header and a file map. The old filecontents are used as if they were one strip. It may not be possible to convert a fragmented file to contiguous file, since the fragmented file may contain mapping for a larger logical space than exist on the media.

   Fragmented files have not been used to store directories, since there is a substantial performance penalty associated with fragmented files, especially if there are many tiny fragments. Fragmented files cannot be used to play back audio or video in real time, owing to the interruptions caused by seek operations between strips.

   If a single byte in the middle of a large fragment is modified, a new fragment must be generated to hold the modified byte and some amount of surrounding information. The old fragment is split into two fragments.  This takes advantage of the cdblock pointer's ability to point to any byte on the media, so that a strip can begin in the middle of a sector. The data of the new fragment and a new file map is written to the media, with one modified and two additional fragment descriptors.

   CDFS will provide primitives sufficient to read and write files with holes in them. Primitives will be written to insert and delete data from within files, changing the position of subsequent bytes in the file, in addition to the traditional read, write and positioning functions.

7.2. Addnames
   Each file or directory has a single location in the hierarchical directory tree. Names in directories can be up to 48 characters long. This permits long and descriptive names for files, and ensures that names from most other systems will fit without exceeding CDFS's length limit. Pathnames can be much longer, since they are a delimited concatenation of directory names and a file name.

   Computer users often view viewed these characteristics as limitations. It is often helpful to have a file or directory appear at more than one location, transforming a tree structured hierarchy into a network. Long names may be descriptive, but they are tedious to type; short names ease manipulation of the file system. Files or directories may have more than one purpose. Having a single name for a file can be very restrictive.

   CDFS addresses these problems with two mechanisms: addnames and links, neither implemented in the prototype implementation. Both constructs appear in other file systems, so their presence in CDFS makes it possible to back up addnames and links from those file systems.

   An addname is an additional name on a file or directory. After a file or directory is created with one name (the primary name), additional names can be added. Names can be deleted, but there must always be at least one name associated with any given file. CDFS programmer's interface calls behave in the same manner regardless of which addname of a file is specified. Renaming an addname does not affect the other addnames of that file. Deleting a file with addnames deletes the file, whether the supplied name was the primary name or an add name.  If the primary name is deleted, an addname is promoted to primary name.  All of a file's addnames are confined to the same directory. A particular name can appear only once inside a directory, and can be associated with exactly one file.

   Addnames are stored in a CDFS directory as entries which have a filetype of Addname. Addnames are resolved by searching the containing directory for a non-addname entry with the same filenumber as the addname. Information such as the file version number and file location is found from the primary name's entry.

   The primary name's directory entry contains a count of its addnames. When the directory is listed, this can be used to allocate storage to hold all the addnames, and the addnames can be filled in on a second pass of the directory array. The directory array is kept sorted in character collating sequence order for binary searching. If an addname count is zero, then that file has no addnames.  We decided not to link the names together with pointers or array indices as this complicates the maintenance of the directory, requires buffering the entire directory to modify it, and provides opportunities for inconsistent directories.

   Only the file_name, modify_time, file_number and file_type directory entry fields of an addname are relevant.

7.3. Soft Links
   A link is an entry in a directory which points to another entry (called the Target) possibly in another directory. CDFS establishes a generalized link concept known of as a Soft Link.

   Soft links consist of an entry in the containing directory and a fileheader, in which the file_info_ structure of the fileheader has been replaced by a soft_link_info_ structure.(see figure FILEHEADER-FIGURE) The soft link Target information consists of two or three parts: a directory number, a character string, and an optional version number. The directory number is the starting point for the resolution of the character string. When the directory number is 1 (the root directory), the soft link corresponds to "symbolic link" found on other systems. The directory number is intended to allow pathnames to be relative either to the root or to the containing directory, however any directory number can be used.

The character string consists of a filename, or a delimited concatenation of
file and directory names. There are two delimiters, which are reserved
non-ASCII characters not permitted in file names. The first delimiter, called
"down," (octal character 0376) indicates the end of a directory name. The
second delimiter, called "up," (octal character 0375) indicates that the next
name is to be resolved in the directory containing the directory or file name
that the up character terminates. If the name is a link, the link target is
first determined recursively. If an "up" occurs at the beginning of the
character string, the directory containing the initial directory is used. If an
"up" occurs immediately following an "up," the directory containing the
currently selected directory is used.

   If a version number is specified as part of the soft link it may be
necessary to search for the requested version once the Target is located.

   Soft links may be created with nonexistent targets.

   Soft links can designate pathnames on other volumes. The volume name is
stored in the property list.

   A second type of link, termed "Firm Links," is currently under
consideration. A Firm Link would specify a directory number and a file number,
rather than to a directory number and a name. It would be unaffected by
renaming the Target.

8. Summary
   The Compact Disk File System (CDFS) has been developed by the MIT Media Lab
to allow for the immediate exploitation of DRAW devices as they become
available. The CDFS was designed to provide the useful capabilities of
multiple-write file systems on write-once media. The CDFS also provides new
functionality as a consequence of the write-once nature of the media.
Specifically, it provides for a complete history of every file and the ability
to recover files after they have been "deleted".

   CDFS stores all information pertaining to a given volume on the volume
itself. No directory information need be stored on associated multiple-write
media, however performance may be enhanced by caching information on
multiple-write device. This permits a file system resident on a CD to be moved
from one site to another by merely moving the disk from one drive to another.

   CDFS is proposed as a standard so that transportability of CDs can transcend
operating systems. Any CD written under any operating system is readable under
any other operating system equipped with CDFS and a compatible drive.  The
standard is a useful superset of all commonly available file systems and not a
subset interchange standard. CDFS stores most information about file contents
and attributes that are maintained by most operating systems including UNIX,
MULTICS, VMS, VM/CMS, MSDOS, TOPS-20, and many others.  This permits, for
example, UNIX hierarchies and VM/CMS minidisks to be copied to the same CD and
accessed using either system.

   CDFS maintains the appearance of a traditional read/write, hierarchical file
system, with directories, files, file property lists and links. The standard
also includes provisions for coherent extension. Certain aspects of file
systems which are inconvenient to model have been omitted, notably permitting
the same file to reside in more than one directory at the same time (such as
"hard linked" files in a UNIX file system).

CDFS uses a "layered" approach to implementation. Specifics such as block
size, block addressing, and device capacity are relegated to a "driver layer"
which need provide a small number of standard primitives to the CDFS. Driver
layers can be written to allow use of the CDFS with any write-once media,
including DRAW digital video disks and punched cards.

   Under normal operation, CDFS never overwrites the contents of a block,
although primitives are provided for the deliberate destruction of data.
Instead, CDFS continually writes new versions of files, directories and other
volume information. Normally, the "current" or most recent version is
automatically selected and updated. Earlier versions of files and directories
can be quickly retrieved.

   CDFS does not require the preallocation of separate data and directory
partitions on the CD. The CD is treated as a stream of blocks which are written
sequentially. Thus, at all times, the CD consists of two regions:  blocks that
have been written and "virgin" blocks that have not been written.  A file
system which preallocated space would either unreasonably restrict the user
(who might have run out of directory space while ample file space remained) or
require complex overflow area management. Procedures for the partitioning of
the CD and the addressing of data (such as music or preallocated files) located
outside of the file system partition but on the same CD have been established.
Anticipated low-cost hardware which cannot write to any addressable block, but
only append to the end of the modified area can be used with the CDFS (although
the operation of destroying data or writing to an alternate partition cannot be
supported on such a drive).

   While CD drives have high transfer rates, they have long seek times.  With
this in mind, the CDFS has been designed to minimize the number of seeks
required to find any file on a volume. Mounting a CD may require up to twenty
seeks to locate the end of the written block stream. Once a CD is mounted, any
file's current version can be located in at most one seek per level of depth in
the directory tree.  An additional seek may be required to read the contents of
the file.

   CDFS stores a substantial amount of redundant information on the disk,
allowing accurate reconstruction of a volume after media failure or partial
update.

   CDFS does not require that files, directories, or other volume information
be written in any particular order on the CD. Very little information need be
cached in a minimum-memory implementation, although such an implementation will
be forced to write these elements in a particular order. This ordering is the
ordering chosen in our implementation. Other orderings might be preferred if
significant buffer memory is available.

9. Notes on CDFS data structures
   The End-Of-Transaction, Directory List and Fileheader structures all contain
three distinct means of validation. These are an identifying constant string, a
checksum and a self-referential pointer. The concurrent use of all three
validating methods minimizes the chance of a mis-identification.  Validation is
only required to detect and recover from media failure.

   All CDFS timestamps are 64 bit numbers representing the number of seconds
(not counting leap seconds), since 12:00 Midnight, January 1, 1901 GMT.

   When a structure containing a checksum is taken as an array of int16
quantities and summed, modulo 65536, the total must be zero for the checksum to be valid. The occurance of checksum fields in the structures allows for the sum to be zero.

   Elements in the directory list are sorted by directory number, root directory (1) first. Directory numbers are the file numbers of directories. File numbers are serial numbers, assigned sequentially.

   Note that the directory list, directory array and filemap array are organized so that they need not fit in memory if memory is a scarce resource (and time is cheap). Fixed size entries were used to facilitate binary searching of these arrays.

9.0.1. Notes on End-Of-Transaction
   Encryption_standard is a site-dependent reserved space to allow for the denotion of encrypted disks. The field is ignored in the Media Lab's CDFS implementation.

   Owners_name is a NULL terminated string. The length of the string may be calculated by subtracting the offset of owners_name in the structure from eot_length.

9.0.2. Notes on Directory List
   The directory list array is stored sorted by directory number.

   The modify_time of a directory is the most recent modification time of any of its contents.

   The contained_bytes of a directory is sum of the sizes, in bytes, of all contained files plus the contained_bytes value for all contained directories. In the case of fragmented files, file size refers to the number of valid data bytes, not the highest valid data address.

   Modify_time and Contained_bytes are calculated when Transactions are closed.

9.0.3. Notes on Directory Structures
   Directory elements are stored in a packed array following the directory_info structure. The directory_info structure is pointed to by the directory's fileheader.

   Directory entries are stored with 48 characters reserved for the entry name. These entries are padded with the NULL character but need not be NULL terminated. The octal characters 000 (NULL), 0376 (down) and 0375 (up) are reserved and may not be part of an entry name.

   Since the header_location for a directory_entry that is itself an entry is meaningless, a location of "0" is used. The location of the contained directory must be found through the directory list.

   Entries are stored sorted alphabetically on file_name for fast lookup.

9.0.4. Notes on Fileheaders
   access_info_offset, backup_info_offset, file_info_offset, site_info_offset and property_list_offset are figured as byte offsets from the start of the fileheader_.

ACCESS_INFO_VERSION #1 defines file access protections as in the UNIX file system.[In the UNIX file system, every file is assumed to have an owner and belong to a group. Access for owner, group and world are denoted by three bit fields in the access_info_.file_access field. Bits 0-2 denote world access, 3-5 denote group and 6-8 denote world. In the case of files, bit 0 denotes permission to execute the file, bit 1 denotes permission to read the file and bit 2 denotes permission to overwrite the file. In the case of directories, bit 0 denotes permission to reference a contained entry in the directory, bit 1 denotes permission to list the directory and bit 2 denotes permission to make a change to the structure of the directory.] Access control is not enforced by CDFS, since anticipated use is by individuals with physical access to the media. Enforcement policy is to be set by the site and implementation.  Most schemes for access control can be mapped onto the UNIX scheme, but others can be implemented. For backup purposes, additional information can be stored in the property list.

Backup_info_.backup_pathname denotes the location of the file referenced by the fileheader when the fileheader is written. If any containing directory is renamed or moved, backup_pathname will be invalid.  filename_offset is the byte-offset of the filename in backup_pathname.

The property list begins with the property_list_info_ structure.  If the offset to this structure is 0, then there is no property list.  A property list consists of a group of two-element ASCII records.  It is used to denote a set of user-definable properties associated with the file.

Properties may be flags, such as "BITSTREAM" or "CLASSIFIED", in which their property_value_len is 0. Values, if specified, must be in ASCII, to eliminate problems associated with multiple binary and floating point representations across processors.

The following properties are at the present time defined:

BITSTREAM      Denotes that the information stored in the file is to be taken
               as a stream of bits and not as a stream of characters. This has
               relevance on systems whose byte size is not 8 bits.

               When BITSTREAM is in effect, the UNIT property has meaning.

PARTITION      Denotes a file whose contents are allocated outside of CDFS.

RECLEN nnn     Record size of the file, expressed in characters.

UNIT           Size of the byte of the machine that created the BITSTREAM. On
               a PDP-10, this would have the value of "36".

VOLUME-NAME    For use with soft links, denotes a link to a file resident on
               another CD.

                        Figure 9-1:   File Type definitions

```
#define FILE_TYPE        1
#define DIRECTORY_TYPE   2
#define SOFT_LINK_TYPE   3
#define FRAGMENTED_TYPE  4
#define FIRM_LINK_TYPE   5
```

```
#define ADDNAME_TYPE    6
```

Figure 9-2:   End Of Transaction structure format

```
typedef struct {
     int32     modulo_of_value;
     int16     bits_in_value;
     int16     pad;
} pointerdef;

#define EOT_ID_STRING_LEN 8
#define EOT_ID_STRING "\237\002\CDFS\255\000"
#define EOT_VERSION 1
#define CDFS_IMPLEMENTATION_ID 1

typedef struct {
     char id_string[ EOT_ID_STRING_LEN ];

     int16 eot_version;
     int16 eot_length;

     cdblock    eot_location;

     int16      eot_checksum;
     int16      CDFS_implementation_id;

     cdblock    current_dir_list;
     cdblock    previous_eot_location;
     cdblock    next_eot_location;

     int64      filesystem_creation_time;
     int32      trans_number;
     int64      trans_start_time;
     int64      trans_end_time;
     int32      files_written_on_trans;
     int32      dirs_written_on_trans;
     int32      next_free_file_number;

     pointerdef pointerdes[16];

     int16      number_of_used_pointerdefs;
     char       encryption_standard[32];
     char       owners_name[ variable ];

} eot_format_ ;

typedef struct {
     int32     modulo_of_value;
     int16     bits_in_value;
     int16     pad;
} pointerdef;
```

Figure 9-3:   Directory List format

```c
#define DL_ID_STRING_LEN 8
#define DL_ID_STRING "\237\001CDFS\250\000"
#define DIR_LIST_VERSION 1

typedef struct {
  char    id_string[ DL_ID_STRING_LEN ];

  int16   dir_list_version;
  int16   dir_list_header_length;
  cdblock dir_list_loc;

  int16   dir_list_checksum;
  int16   pad;

  cdblock prev_dir_list;

  int32   dir_list_entry_count;
} dir_list_;

typedef struct {
      int32    dir_number;
      cdblock  header_location;
      int32    containing_dir;
      int64    modify_time;
      int64    contained_bytes;
      int16    header_size;
      int16    pad;
} list_element_;
```

Figure 9-4:   Directory format

```c
#define DIRECTORY_INFO_VERSION_ 1
typedef struct {
        int32           directory_info_version;
        int32           directory_info_length;

        int32           directory_entries;
        int32           directory_entry_size;
        } directory_info_;

#define MAX_COMP_LEN 48
typedef struct {
      char   file_name[ MAX_COMP_LEN ];
      cdblock header_location;
      int64  modify_time;
      int32  file_number;
      int32  file_size;
      int32  file_version;
      int16  file_type;
      int16  header_size;
      int16  addname_count;
      int16  pad;
  } dir_contents_;
```

Figure 9-5: Fileheader structure definition

```
#define FH_ID_STRING_LEN 8
#define FH_ID_STRING "\237\001\CDFS\255\000"

#define HEADER_VERSION 1
typedef struct  {
  char    id_string[ FH_ID_STRING_LEN ];
  int16   header_version;
  int16   header_length;

  int16   header_checksum;
  int16   fileheader_length;

  cdblock fileheader_location;

  int32   file_number;
  int16   file_type;

  int16   access_info_offset;
  int16   backup_info_offset;
  int16   file_info_offset;
  int16   site_info_offset;
  int16   property_list_offset;

} fileheader_;

#define ACCESS_INFO_VERSION 1
#define GROUPLEN 32
#define OWNERLEN 32
typedef struct {
  int16   access_info_version;
  int16   access_info_length;

  char    file_owner[OWNERLEN];
  char    file_group[GROUPLEN];
  int16   file_access;
} access_info_ ;

/* Directory component delimiter in backup_pathname */
#define DOWN_DIR_CHAR 0376
#define UP_DIR_CHAR   0375

#define BACKUP_INFO_VERSION 1
typedef struct  {
  int16   backup_info_version;
  int16   backup_info_length;

  int32   containing_directory_number;
  cdblock previous_version_location;
  cdblock previous_eot_location;
  int16   filename_offset;
  int16   previous_version_header_size;
  char    backup_pathname[ variable ];

} backup_info_;

/*
```

```
   Note that backup_pathname will be wrong if any containing directory is
subsequently renamed.
 */

#define FILE_INFO_VERSION 1
typedef struct {
  int16   file_info_version;
  int16   file_info_length;

  cdblock file_location;
  int32   file_length;
  int64   write_time;
  int64   creation_time;
  int32   file_version_number;
} file_info_ ;

/* If block is a soft link, use soft_link_info_
 * to decode file_info */

#define SOFT_LINK_VERSION 1
typedef struct {
  int16   soft_link_info_version;
  int16   soft_link_info_length;

  int64   creation_time;
  int32   target_dir;
  int32   target_version;
  char    target_name[ variable ];
} soft_link_info_ ;

#define SITE_INFO_VERSION 1
typedef struct {
  int16   site_info_version;
  int16   site_info_length;

  char    opsys[16];
  char    opsys_version[16];
  char    site_name[ variable ];
} site_info_ ;

#define PROPERTY_LIST_VERSION 1
typedef struct {
  int32   property_list_version_;
  int16   property_list_length;

  int16   property_list_entries;

} property_list_info_;

typedef struct {
  int16   property_name_len;
  int16   property_value_len;
  char    property_name[ variable ];
  char    property_value[ variable ];
} property_list_record;
```

Figure 9-6:   File Map format for fragmented files

```
#define STRIP_INFO_VERSION 1
typedef struct {
  int32    strip_info_version;
  int32    strip_info_length;

  int32    strip_count;
} strip_info_;

typedef struct {
  cdblock loc;         /* location of first byte in strip. */
  int32   valid_chars; /* number of valid bytes in strip */
  int32   ordinal;     /* byte offset in logical file of strip */
} fragmented_des;
```